\documentclass[usenatbib,useAMS]{mn2e}

\usepackage{natbib}
\usepackage{graphicx,epsfig,float}
\usepackage{latexsym,amssymb}
\usepackage[fleqn]{amsmath}
\usepackage{color}
\usepackage{txfonts}
\usepackage{verbatim}
\usepackage{times}

\newcommand{\gcrit}{\gamma_{\mathrm{crit}}}
\newcommand{\refeq}[1]{(\ref{#1})}

\title[Gravitational Collapse in One Dimension]
      {Gravitational Collapse in One Dimension}

\author[Schulz et al.]{%
  A.E. Schulz$^1$\thanks{\texttt{aschulz@lanl.gov}},
  Walter Dehnen$^2$,
  Gerard Jungman$^1$,
  Scott Tremaine$^3$
  \\
  $^1$Los Alamos National Laboratory, P.O.~Box 1663, MS-B227,
  Los Alamos, NM 87545, USA 
  \\
  $^2$University of Leicester, University Road, Leicester LE1~7RH, UK
  \\
  $^3$Institute for Advanced Study, Einstein Drive, Princeton NJ 08540, USA 
}

\begin{document}

\date{\today}

\maketitle 

\begin{abstract}
We simulate the evolution of one-dimensional gravitating collisionless systems from
non-equilibrium initial conditions, similar to the conditions that lead to
the formation of dark-matter halos in three dimensions.
As in the case of 3D halo formation we find that initially cold, nearly homogeneous particle distributions 
collapse to approach a final equilibrium state with a
universal density profile.
At small radii, this attractor exhibits a power-law behavior in density,
\(\rho(x) \propto |x|^{-\gcrit}\), \(\gcrit\simeq0.47\), slightly but
significantly shallower than the value $\gamma=1/2$ suggested previously.
This state develops from the initial conditions 
through a process of phase mixing and violent relaxation.  This process 
preserves the energy ranks of particles.
By warming the initial conditions, we illustrate a cross-over from this
power-law final state to a final state containing a homogeneous core.
We further show that inhomogeneous but cold power-law initial conditions, with
initial exponent \(\gamma_i > \gcrit\), do not evolve toward the attractor
but  reach a final state that retains their original power-law behavior in the interior of the
profile, indicating a bifurcation in the final state as a function of the initial exponent.
Our results rely on a high-fidelity event-driven simulation technique.
\end{abstract}


\section{Introduction}\label{sec:intro}

The study of gravitating many-body systems is central to a wide variety of
problems in astrophysics and cosmology. Perhaps the most important of these
problems is the development of structure in the non-baryonic dark matter that
dominates the mass budget of the universe. Dark matter is believed to be
``cold,'' that is, the velocity dispersion in the initial state is negligible.
Such an initial state is gravitationally unstable, and subsequent evolution leads to
complex three-dimensional structures such as halos, sheets, filaments
and streams. Halos are the most recognizable structures.
Structures such as filaments or streams inside halos are identified as coherent
patterns in phase space, less readily visible but of significant importance.
The phase-space structure of collapsed halos is important 
mainly because of its relevance for estimating rates and
uncertainties for direct detection of particle dark matter
\citep{2011MNRAS.415.2475M,2011MNRAS.413.1419V}
and for indirect detection from dark-matter annihilation near the Galactic center
\citep{2007PhRvD..76h3012H,2010ApJ...717..825D} or in halo substructure
\citep{2008Natur.454..735D,2011arXiv1108.3546T}.
Currently, the dynamics of these systems can be studied only with
large numerical $N$-body simulations.

Cosmological $N$-body simulations reveal a number of common features of halos
that so far have resisted analytic explanation. The most striking feature is
that the spherically averaged density profiles of collapsed halos appear to have
the same form, apart from scaling factors for the total mass and radius. The common
profile can be approximated by the \cite*{1996ApJ...462..563N} or NFW formula,
$\rho(r)\propto r^{-1}(r+a)^{-2}$ or the \cite{1972TarOT..36....3E} formula
$\rho(r)\propto \exp[-(r/a)^\alpha]$. A more detailed and better motivated
analysis of shape universality is given by \cite{2005MNRAS.363.1057D}.
See \cite{2006AJ....132.2685M} for a comparison of models.
The existence of a ``universal'' final state for equilibrium
halos is reminiscent of Lynden-Bell's (1967)\nocite{1967MNRAS.136..101L}
argument that chaotic changes in energy caused by the rapidly varying
gravitational potential during collapse lead to an equilibrium state that can
be determined by statistical mechanics. Lynden-Bell called this process
``violent relaxation'' and we shall use this term as well, although without
the implication that the final equilibrium state can be computed by methods of
statistical mechanics or that the changes in energy are chaotic. 

By their nature, $N$-body simulations of fully three-dimensional systems
probe only a limited range of length scales.
Systems of lower dimensionality, or having special symmetries, provide laboratories
in which to explore many-body dynamics with far higher phase-space resolution than
is otherwise possible, and studies of these simpler systems can suggest
explanations for behavior seen in the three-dimensional case, such as
the emergence of a universal halo profile (see \citealt{2010arXiv1010.2539D}
and references therein).
Furthermore, simulations of reduced systems may exhibit some of the
same systematic problems as simulations of three-dimensional dynamics,
providing an opportunity for high-fidelity tests of the overall methodology.
Recently, \cite{2004MNRAS.350..939B} argued that simulations of one-dimensional
systems indicate that discretization errors affect the central
structure of simulated collapsed objects. He also observed
that violent relaxation led to a power-law density profile,
\(\rho(x)\propto |x|^{-\gamma}\) with \(\gamma\sim1/2\), over three orders of magnitude
in radius, although this was not the first time
that this type of collapse had been investigated \citep{1063-7869-38-7-A02}.
Although he examined only one set of initial conditions, Binney conjectured that this power law was
universal over a wide variety of ``cold'' initial conditions. An understanding of this
scaling could illuminate the origin of the central density cusps found in simulations
of three-dimensional collapse.

In this paper, we report simulations of the formation of cold
dark matter halos having ``slab'' symmetry, in which the density is independent of
two of the three spatial Cartesian coordinates. This system is equivalent to a
one-dimensional system of gravitating particles, as studied by Binney and others.
Our simulations improve the phase-space resolution of previous work by more than
an order of magnitude, and explore a much wider range of initial conditions. 
We find strong evidence that cold collapse leads to final equilibria that are 
close to a universal power-law density profile, with an exponent less than \(1/2\).

When the initial conditions are slightly ``warmed'', we observe a cross-over
from the power-law behavior of cold systems to the cored behavior displayed
by warm systems. This dichotomy raises the question of the meaning of ``coldness'',
which we discuss and relate to our simulation results.
We have also examined inhomogeneous initial states, with power-law density profiles,
and observed an apparent bifurcation in the dynamical behavior of the system as a
function of the initial power-law exponent.
Initial states with power-law density
profiles shallower than \(|x|^{-\gcrit}\) with \(\gcrit \simeq 0.47\) evolve to a
universal profile \(|x|^{-\gcrit}\); initial states with profiles steeper than
\(|x|^{-\gcrit}\) display a persistence of their initial power law. Therefore,
the \(|x|^{-\gcrit}\) profile represents a type of dynamical attractor,
accessed by ``shallow" initial states.  ``Steep" initial states 
are not attracted to the critical scaling solution.

In section \ref{sec:odg} we explain the basic setup for one-dimensional
gravitational collapse and discuss the elementary properties of stationary
self-similar solutions.
In section \ref{sec:sim} we describe the algorithm used to perform the
simulations in this work.
Section \ref{sec:res} details our numerical results on the density profile,
the particle energies, and the scaling properties of the final state.
Section \ref{sec:conclusions} summarizes our findings.


\section{One-dimensional gravitating \(\mathbf{N}\)-body systems}
\label{sec:odg}

One-dimensional gravitating (1DG) systems provide the simplest possible
framework in which to investigate the nature and outcome of gravitational
collapse. Violent relaxation and phase mixing \citep{2008gady.book.....B,1998ApJ...500..120K}
in such systems have long been a subject of computational study
\citep{1969MNRAS.146..161C,1971Ap&SS..13..397L}.
Simulations of 1DG structure formation with cold dark matter and cosmological
initial conditions date back to \cite{1980MNRAS.192..321D}.
Because of our interest in cosmological systems with very large
numbers of particles, we focus on the collisionless limit where
the number of particles tends to infinity, holding total mass and energy fixed,
and keeping time bounded \citep{1977CMaPh..56..101B}.
The mean-field dynamics of an isolated system described by this limit satisfies the collisionless
Boltzmann equation (CBE), with potential and forces determined by the Poisson equation;
\begin{align}
& \frac{\partial f}{\partial t} + 
  \upsilon\,\frac{\partial f}{\partial x} -
  \frac{\partial\Phi}{\partial x}\,\frac{\partial f}{\partial \upsilon} =0 
  \label{eq:cbeq}\\
  &\frac{d^2\Phi}{d x^2}=4\pi G\rho=4 \pi G m_p \int f\, \mathrm{d}\upsilon;
  \label{eq:pois}
\end{align}
Here \(x\) and \(\upsilon\) are position and velocity coordinates in the two-dimensional
one-particle phase space, \(f(x,\upsilon,t)\) is the particle number density in this space,
\(\rho(x,t)\) is the mass density, \(\Phi(x,t)\) is the gravitational potential, and \(m_p\) is the 
particle mass.
In this limit, two-body interactions are negligible. For collisional 1DG systems
see \cite{1971Ap&SS..14...56R},\cite{PhysRevE.56.2429} and \cite{PhysRevE.80.041108}.

If the potential is stationary, the specific energy $E=\frac{1}{2}\upsilon^2+\Phi(x)$ is an
integral of the motion. By Jeans' theorem, stationary solutions of the
collisionless Boltzmann equation can only depend on the integrals of motion
so $f(x,\upsilon)=f(E)$. Then the Poisson equation (\ref{eq:pois}) becomes
\begin{equation}
  \frac{\mathrm{d}^2\Phi}{\mathrm{d}x^2}=8\pi G m_p\int_\Phi^\infty
  \frac{\mathrm{d}E\,f(E)}{(2E-2\Phi)^{1/2}}.
\end{equation}
We may multiply this by $d\Phi/dx$ and integrate once to obtain
\begin{equation}
  \left(
  \frac{\mathrm{d}\Phi}{\mathrm{d}x}\right)^2=w(\Phi)
  \quad\text{where}\quad
  \frac{\mathrm{d}w}{\mathrm{d}\Phi}=8\pi G\int_\Phi^\infty
  \frac{\mathrm{d}E\,f(E)}{(E-\Phi)^{1/2}}.
\end{equation}
Re-writing this equation in the form 
\begin{equation}
  \frac{\mathrm{d}x}{\mathrm{d}\Phi}=\pm\frac{1}{\sqrt{w(\Phi)}},
\end{equation}
we see that all stationary solutions must have left-right symmetry
about some point $x_0$ which we may choose to be the origin, i.e.,
$\Phi(x)$ and $\rho(x)$ are even functions of $x$.

Among all stationary solutions, a family of self-similar solutions can be
constructed by choosing the phase-space distribution to be a power law
in energy,
\begin{align}
  f(E) = E_0^{-1/2} (E/E_0)^{-p}.
  \label{eq:df}
\end{align}
 We have chosen to make the position coordinate dimensionless, and factors of 
 $E_0$ have been inserted so that $f$ has dimensions of inverse velocity.  
 Computing the density from equation \refeq{eq:pois} gives
\begin{align}
  \rho(x)=
  \sqrt{2}\,B\left(\tfrac{1}{2},p-\tfrac{1}{2}\right)\,
  m_p (\Phi/E_0)^{1/2-p},\quad p>\tfrac{1}{2},
\end{align}
where $B(a,b)$ is the beta function. Choosing a center of symmetry where the
potential is set to zero, we write the potential, density, and enclosed mass
in the form
\begin{align} \label{eqn:rhophi}
  \rho(x)=\rho_0|x|^{-\gamma},\quad
  m(x)=m_0|x|^{1-\gamma}, \quad
    \Phi(x)=\Phi_0|x|^{2-\gamma} + \Phi(0), 
\end{align}
where 
\begin{align}
  \gamma=\frac{4p-2}{1+2p}.
\end{align}
The constraint \(p>\frac{1}{2}\) implies \(\gamma > 0\), and the requirement
that any compact interval containing the origin has finite mass implies
\(\gamma < 1\) or \(p<\frac{3}{2}\).  The relation between $E_0$,
$\Phi_0$, and $\rho_0$ is  straightforward to derive but is not needed here. 

Self-similar solutions can also be described in terms of the action,
\begin{align}\label{eqn:actiondefinition}
  J(E) = \frac{1}{2\pi} \oint \upsilon\, \mathrm{d}x 
  \,=\frac{2}{\pi} \int_0^{x_{\rm max}} \sqrt{2E - 2\Phi(x)}
  \;\mathrm{d}x,
\end{align}
where \(E=\Phi(x_{\max})\), and \(x_{\max}\) is the maximum excursion of the orbit from the center.
For a power-law stationary state, the relation between energy and action is
\begin{align}
  J(E) = \frac{2^{3/2}E^{1/2}}{\pi(2-\gamma)} 
  \left(\frac{E}{\Phi_0}\right)^{1/(2-\gamma)}
  B\left(\tfrac{1}{2-\gamma},\tfrac{3}{2}\right).
\end{align}
Since $J\propto E^{(4-\gamma)/(4-2\gamma)}$, the phase-space distribution function
(\ref{eq:df}) has the power-law form
\begin{align}\label{eqn:action}
  f(J) &\propto J^{-(2+\gamma)/(4-\gamma)} \propto J^{-4p/(3+2p)}.
\end{align}
Because of its relation to energy, the maximum excursion of an orbit
is also a useful surrogate for investigating the scaling of the potential.

The emergence of a self-similar final state from cold collapse in one dimension
was observed by \cite{1063-7869-38-7-A02} and \cite{2004MNRAS.350..939B}.
An approximate analytic theory for the self-similar state was presented in 
\cite{1063-7869-38-7-A02}, with
an estimated exponent \(\gamma \simeq 4/7\); simulations described
in that work were claimed to be consistent with this approximate exponent
but few details were given. Their theoretical estimate is unlikely to be accurate because it 
is based on the assumption that adiabatic evolution begins at the transition 
from one to three streams; we have observed empirically 
that the actions evolve after this time.  
The exponent observed in \cite{2004MNRAS.350..939B}
was estimated there to be \(\gamma \simeq 1/2\).

The meaning of coldness in the initial states requires some discussion.
Intuitively, a state is cold if the velocity dispersion is small
compared to the average potential. In the continuum limit,
a perfectly cold state corresponds to a phase-space
density that is supported on a smooth curve in the \((x,\upsilon)\) plane,
with only one velocity at any location.  
In the context of a particle simulation, a state is cold if it well-approximates
such a continuum state. 
These statements fall short of being a precise definition,
but the meaning of coldness will be further examined in the context of our simulation
results, described below.  A cold system in the continuum limit remains 
a sub-manifold in phase space indefinitely.  However, a particle realization of a cold initial condition 
could depart from the behavior 
of the continuum model, either due to particles
leaving the lower dimensional surface or by under-resolving it.  These effects can
artificially heat the system by introducing additional velocity dispersion to a coherant stream 
of particles at a given location.  We will see 
that the low dimensionality of the phase-space surface 
is very well preserved in the particle realizations, 
until late times when discreteness effects grow
to importance.

Any coordinate parametrization for a curve in the \((x,\upsilon)\) plane provides a Lagrangian
coordinate for particles along the curve. Particle number or rank along the curve is one such
coordinate. Up to an irrelevant numerical factor, and allowing for continuation
to negative values, the "mass enclosed on the curve" is equivalent to this
particle number. In a self-similar state, this Lagrangian mass coordinate
satisfies a power-law relation of the form
\begin{align}\label{eqn:Jvsmu}
\mu(J) \propto \int_0^J f(J')\, dJ' \propto J^{(2-2\gamma)/(4-\gamma)},
\end{align}
using the expression from equation \refeq{eqn:action}.  Note that 
the expression in equation \refeq{eqn:action} is for equilibrium systems that are 
not necessarily cold, and that in this context is only valid at times when
the system is very tightly wound and the orbital phases are mixed. 
Identifying \(\mu\) with initial particle rank, this relation provides another
avenue for investigation of the scaling properties of the final stationary state,
as long as this identification remains valid. The identification
becomes invalid when \(J\) is no longer a monotonic function of particle rank.


\section{Algorithm}
\label{sec:sim}

The system studied here consists of \(N\) parallel sheets, each of
surface density \(\Sigma/N\), interacting only through their mutual gravitational force;
we shall call these sheets ``particles'' and treat them as an effective one-dimensional
gravitating system. When two particles collide, they can either be passed through
each other with no change in velocity or scattered elastically; the resulting dynamics
is the same in either case, up to a possible relabelling of particles. For definiteness,
we choose to let particles pass through each other at collisions. With this choice,
the initial positional rank of particles in a cold state remains a suitable Lagrangian
coordinate throughout the evolution.

The potential energy between two particles is
\begin{equation} \label{eq:pot}
  \Phi_{i\!j} = 2\pi G\,(\Sigma/N)^2\,|x_i-x_j|,
\end{equation}
such that the force on particle \(i\) due to particle \(j\) is
\begin{equation} \label{eq:force}
  F_{\!i\!j} = -\partial\Phi_{i\!j}/\partial x_i = -2\pi G\,(\Sigma/N)^2 \mathrm{sign}(x_i-x_j).
\end{equation}
The resulting net acceleration of particle \(i\) due to all other particles is
dependent only on its positional rank, \(k_i\), with respect to the median position.
Particles to the right of the median position have \(k_i>0\)
and those to the left have \(k_i<0\). 
Simulations with odd particle number have a particle at $x=0$ and integer values of
rank.  Simulations with an even number of particles have half-integer rank. 
Only particles \(j\) satisfying \(|k_j|\le|k_i|\) contribute
to the net force on the particle \(i\); forces from \(|k_j| > |k_i|\) are balanced right and left.
In the rest of the paper, and in all our reported simulations, we choose units
such that \(2\pi G = 1\) and \(\Sigma = 1\).

The form of the force law implies a great simplification for simulations of this system.
The acceleration remains constant as long as the particle's
rank \(k_i\) remains unchanged. Between particle crossings, the orbits \(x_i(t)\) are
quadratic in time \(t\), and the time of the next crossing of particles \(i\) and
\(j\) can be obtained by solving the quadratic equation \(x_i(t) = x_j(t)\).
Therefore, a numerical simulation can follow the exact evolution of the system,
with errors arising only from roundoff in the solution of the quadratic equation
and in the machine representation of phase-space coordinates.

Conceptually, the calculation consists of the following
steps. Compute all crossing times between each particle and its two neighbors
on either side; find the earliest crossing time; advance to that time and
update all particle positions; swap the crossing particles and adjust their
accelerations; recurse. However, since the acceleration of non-crossing particles is
unaffected by the crossing, it is not necessary to update the positions or re-calculate
crossing times for non-crossing pairs or their neighbors. 
Nor is it necessary to advance non-crossing particles
in time, except when the total particle state is synchronized for output of a data snapshot.
Such simulations are event-driven rather than driven synchronously
in time. The events of the simulation are particle collisions. Similar methods 
for computing the dynamics of one dimensional gravitating systems have been
developed over the last two decades, greatly improving the reach of simulations
\citep{1990A&A...228..344M,PhysRevE.56.2429,2003JCoPh.186..697N}.

One possible code implementation uses a heap data
structure, which stores particle pairs sorted according to their crossing
time \citep{2003JCoPh.186..697N}. This reduces the cost of locating the next
crossing to \(\mathcal{O}(\log N)\), which for very large \(N\) dominates the
operation count, since all other operations per crossing are \(\mathcal{O}(1)\).
However, with this approach it is not straightforward to account for changes
in crossing times between particles \(i\) and \(i+1\) just crossed and their
neighbours \(i-1\) and \(i+2\), since the heap position of these adjacent pairs is
unknown. We developed two different strategies, implemented in two completely
independent codes, to overcome this problem and still preserve the
\(\mathcal{O}(\log N)\) operation count for finding the next crossing.
The first method uses a search tree instead of a heap.  
A further speed-up is achieved by integrating positional
offsets between neighboring particles instead of the particles
themselves.
The second method allows
the heap to contain duplicate collision sets, marked to indicate age of the sets.
To suppress the accumulation of roundoff error, we found it important to
use 80-bit floating-point arithmetic (64-bit mantissa). Simulations run with
ordinary double precision arithmetic (53-bit mantissa) show clear signs of roundoff errors,
starting with a loss of local phase-coherence in simulations from cold initial conditions.

All simulations presented in this paper explicitly enforce left-right
symmetry about \(x=0\), implemented using a mirror boundary condition,
with the particle state reflected about \(x=0\). This effectively doubles the resolution;
the numbers \(N\) reported in the sections below reflect the total number, including mirror
particles. By running simulations without an imposed symmetry, we verified that
our conclusions do not depend on this enforced symmetry nor on the symmetry
of the initial conditions.


\section{Results} \label{sec:res}

\subsection{Description of the Collapse}

\begin{figure}
  \begin{center}
    \resizebox{3.6 in}{!}{\includegraphics{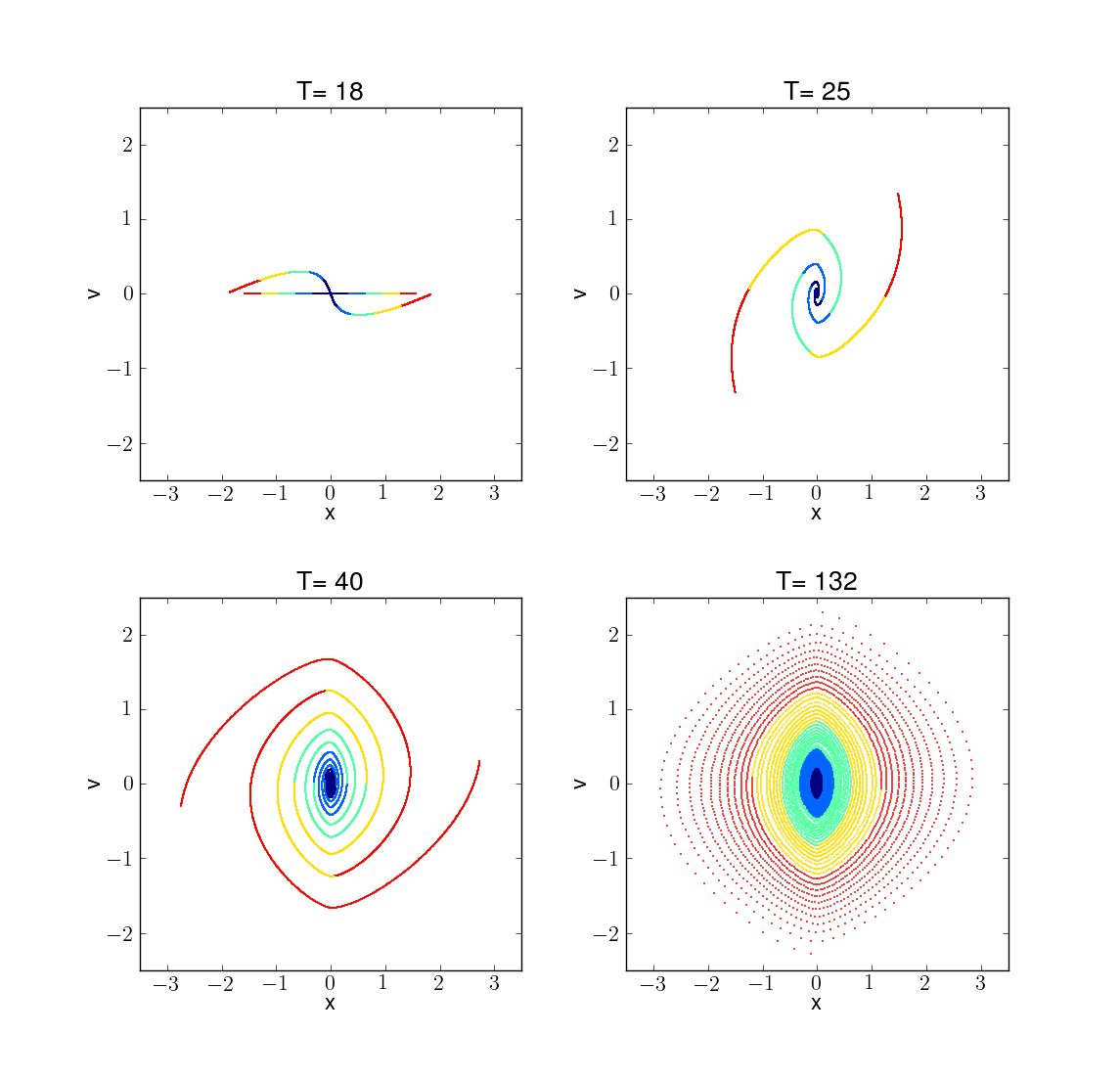}}
  \end{center}
  \caption{The evolution of an initially cold system of \(10^4\) particles with perturbed 
  initial velocities.
  The simulation is described at the beginning of section \ref{sec:results:cold}.
  The spiral locus of the particles winds up as time elapses.  The 
  particle colors are arranged in a spectrum that encode the initial particle positions.  
  Blue particles began the simulation closest to the center \(x=0\), while red ones began farthest 
  from the center. The upper left panel shows two times at and near the start of the simulation.}
\label{fig:evsimple}
\end{figure}

The simplest cold initialization places particles on a smooth curve very near to
but slightly displaced from the \(x\)-axis in the one-particle phase space. Placing
particles with equal spacing in \(x\) gives a homogeneous initial density, and
the velocities can be assigned as a smooth function of \(x\).
It can be shown analytically that initial conditions such as the Hubble flow 
with $\upsilon \propto x$ result in an 
endless repetition of homogeneous collapse and expansion, with no phase mixing
or energy transfer among the particles.  Initial velocities defined with a smoothly varying 
non-linear function of $x$, however, will generate a particle locus that forms an
ever-tightening spiral curve in the \((x,\upsilon)\) plane.  This process is illustrated in 
Fig.  \ref{fig:evsimple}, which shows
four snapshots of the phase space evolution of our fiducial simulation, described
below. We refer to the different arms of the phase space spiral at a given position
as streams.  The particles are divided into five quintiles in initial position, and 
these are color coded in Fig.  \ref{fig:evsimple}, in a spectrum ranging from
blue to red. Blue particles began closest to the center \(|x|=0\), and red particles
began farthest from the center. The density profile of the collapsed object, to be 
discussed later in detail, is the projection of the phase-space curve onto the $x$
axis.  Vertical portions of the phase-space curve in projection create caustic structures
in the density profile (examples can be seen in Fig.  \ref{fig:evol}).

Early in the evolution, rapid changes in the gravitational potential cause energy transfer among the
particles, leading to violent relaxation \citep{1967MNRAS.136..101L}. Note that
the orbital specific energies of particles, \(E_k=\frac{1}{2}\upsilon_k^2+\Phi(x_k,t)\), do not
sum to the conserved total energy, \(E_{\mathrm{tot}}=\frac{1}{2} \sum_k\big[\upsilon_k^2+\Phi(x_k,t)\big]\),
so the energy changes of particles need not sum to zero.
During this period, the potential oscillates in time, with oscillations damping as
the particles approach their final relaxed state.
Violent relaxation ends when the particle locus is tightly wound
and the gravitational potential is no longer changing significantly.

Although the gravitational potential approaches a stationary state, the spiral curve
describing the locus of the particles continues to wind up. In the limit where the
state is described by a continuous curve in the \((x,\upsilon)\)-plane,
this spiral will continue winding indefinitely. In the simulated system,
with a finite number of particles, the particle density along the spiral eventually
decreases to the point where the curve is no longer identifiable; at this
point, discreteness effects have erased any visual remnant of the phase coherence
necessary to define such an approximate curve.

These simple observations are directly related to the meaning of convergence in the
collisionless limit. In order to make contact with the stationary solutions of the
collisionless Boltzmann equation (\ref{eq:cbeq}), we must understand the nature of
this convergence as \(N\rightarrow\infty\). A state supported on a smooth curve
in phase space formally has an infinite number of particles, but it never becomes
stationary in a pointwise sense. Rather, in the collisionless limit, these states converge
only weakly, as measures, to stationary solutions of the CBE \citep{1977CMaPh..56..101B}.
Therefore, the winding spiral locus can be identified with a stationary solution
only through a process of coarse-graining.  Coarse graining need not be restricted
to averaging over phase space coordinates, but could also be implemented by averaging 
over orbital phase.  For example, the $J$ dependence in equation \refeq{eqn:Jvsmu} 
relies on such a coarse graining.

\subsection{Cold initial conditions with homogeneous density}
\label{sec:results:cold}
\subsubsection{The fiducial simulation} \label{sec:fidsim}
To obtain the density profile for the final state of cold collapse, we ran
a suite of simulations, with particle numbers clustering around \(N = 10^4\).
The choice of \(N\) in these simulations is discussed below.
For our fiducial simulation, we chose \(2\pi G = 1\) and set the total mass
to one, \(\Sigma = 1\), implying a system of units where
typical lengths and dynamical times are of order unity.
The particles are initially equally spaced over
the interval \([-\pi/2,\pi/2]\), and the initial velocities are set by
\begin{align} \label{eqn:initvel}
  \upsilon_i(x_i) = -V_0 \sin(x_i),
\end{align}
with the scaling factor \(V_0=0.001\). This state represents an initially
overdense region near turnaround. The evolution
of this state leads to a single collapsed object with a stationary
profile.  The simulations were run to time $t=700$, although several
results are reported at time $t=200$.  At time $700$ ($200$), the outermost 
particles have executed roughly $150$ ($40$) orbits, with inner particles executing 
orders of magnitude more.  The simulation at these times has evaluated 
approximately $1.24 \times 10^{10}$ ($3.5 \times 10^9$) particle crossings.

\begin{figure}
  \begin{center}
    \resizebox{3.6 in}{!}{\includegraphics{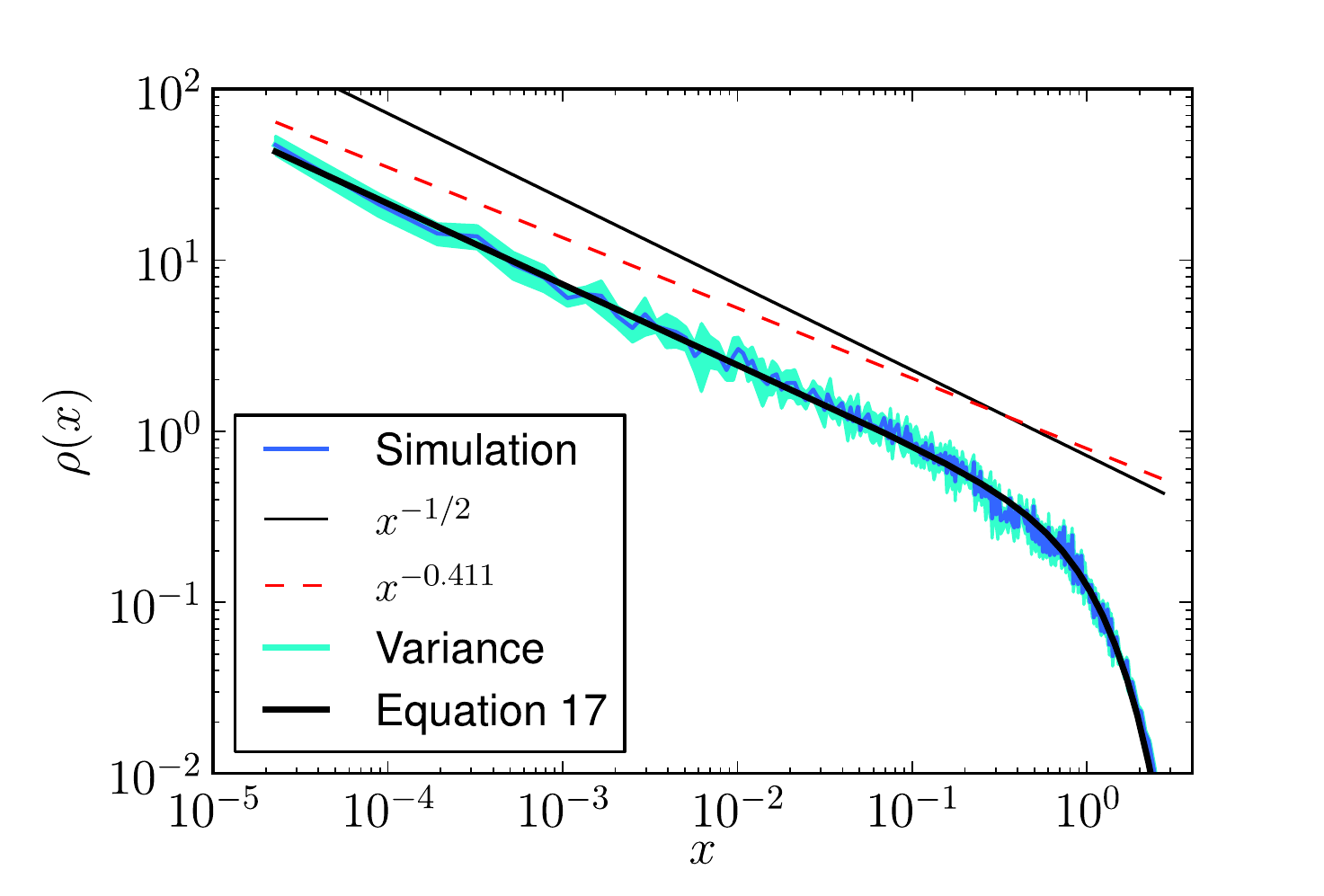}}
  \end{center}
  \caption{The density profile at a time \(t=200\) for a suite of
    cold simulations, with initial conditions as described at the start of
    section \ref{sec:results:cold}.
    The blue line is the density profile averaged over six simulations
    with slightly different particle numbers.
    The shaded region is the error on the mean as determined
    from the variance of the six simulations. The thin black line shows a
    power law \(x^{-1/2}\). The red dashed line shows the shallower power law
    obtained by fitting the relation \(J(\mu)\) of equation \refeq{eqn:Jvsmu}.
    The thick black line shows the model in equation \refeq{eqn:model}.}
\label{fig:rhof1}
\end{figure}

Fig.  \ref{fig:rhof1} shows the density profile of the equilibrated system.
The number density is computed using bins of variable width, set down so that
there are approximately 20 particles per bin, and the abscissae for these points are placed at
the geometric mean of the bin edges.  The curve is the average of six different
simulations.  As is clear from Fig.  \ref{fig:rhof1}, the final state is self-similar over
four orders of magnitude in distance from the center. The fine-scale
structure in the average profile has two easily identifiable sources. First, the
pointwise density profile for each simulation is dominated by caustics, the heights
of which are regulated by the discreteness of the particle representation. Second,
approximating the continuous density distribution along a curve in phase space 
by discrete particles leads to fluctuations in density because of counting statistics. 
The caustic structure is easily identifiable at early
times, but at the time represented in the figure, the discreteness of
particle sampling makes it impossible to identify individual caustics;
in all likelihood there are multiple caustics per bin.

The density estimator for a single simulation approximates 
the continuum density profile but suffers from the fact that the bin spacing 
and the particle phases interact to produce slight (discretized) variations in the
measured density in the bin. Roundoff error may also be responsible for 
occasionally pushing particles over the bin boundaries, a form of discretized
phase error. To explore the impact of discretization and roundoff error, we have repeated
the simulation with slightly different values of \(N\); the six runs used
to produce Fig.  \ref{fig:rhof1} have \(N=10006,\,10012,\,10018,\,10024,\,10030\) and \(10036\).
The bin positions are the same for all six simulations, allowing us to quantify the sensitivity
of the density estimator to the types of phase errors probed by these different choices.
The mean density obtained from these six simulations is shown in Fig.  \ref{fig:rhof1}
as a jagged blue line. The uncertainty in the mean, estimated from the
variance of the six runs, is shown as a shaded region. These ``errors'' are generally
somewhat smaller than Poisson in the power-law region.

The fluctuations in the estimated density should not be interpreted as Gaussian random fluctuations.
An Anderson-Darling test run on the 6 values of the density in each bin suggests 
that only a small fraction of the bins, less than 15\%, are consistent with having been 
drawn from a Gaussian distribution.  A histogram of the skewness of the 6 
density measurements in each bin shows no net skewness; however, most 
of the bins show significant negative kurtosis (rounded peak with very short tails). 
Since fluctuations are principally generated by a small number of
particles shifting across a single bin boundary due to phase differences, we would expect 
particle excursions to be at most one bin division and the tails of the distribution 
to be extremely small.  We have checked that the
distributions of skewness  and kurtosis do not evolve as a function of distance; the inner
decades display the  same skewness and kurtosis properties as the outer decades. Therefore,
we feel confident that we can use the measured variances as weights when
fitting a power law of the form \(\rho(x) =A |x|^{-\gamma}\), without systematically over
or under weighting any region of the data. However, it would be unwise to propagate errors from 
\(\rho(x)\) to errors on the fit parameters \(A\) and \(\gamma\), because such errors 
would be very difficult to interpret.  For this reason, we use the bootstrap technique
to generate a large number of mock datasets, with replacement.  We use a 
least-squares procedure to fit the parameters for each of the mock datasets, 
and report the mean and variance of these fits to the synthetic data. 

Using this method we have fit a power law \(\rho(x) \propto |x|^{-\gamma}\) to the 
portion of the data in the range \(10^{-3} < |x| < 10^{-1}\). We find \(\gamma=0.472\pm0.001\)
with a reduced \(\chi^2_{\mathrm{red}}=0.339\), computed using the standard deviations
within each bin. The small value of the reduced $\chi^2$ compared to the expected
value $\chi_{\rm red}^2=1$ arises from the negative kurtosis discussed earlier. 
The solid black line in Fig. ~\ref{fig:rhof1} is obtained by multiplying this fit with a smooth
truncation function (reminiscent of \citealt{1972TarOT..36....3E} models)
\begin{align}
  \label{eqn:model}
  \rho_{\mathrm{model}}(x) \propto |x|^{-\gamma}\,
  \exp\left(-\left(|x|/x_0\right)^{2-\gamma}\right),
\end{align}
but the truncation of the power law in the outskirts is not the focus of our study.
The fiducial simulation shows that the density follows a power law over at
least four orders of magnitude in distance from the center, 
the extent of which is only limited by our finite
resolution. This confirms the results of \cite{2004MNRAS.350..939B} with many
more particles (301 vs.\ 40000 in our highest N simulation). 
However, the slope \(\gamma=0.472\pm0.001\) differs slightly but significantly
from the value \(1/2\) suggested by Binney, shown in Fig. \ref{fig:rhof1} for comparison.
There is a small systematic trend in the fit as a function of
the range of data being fit; decreasing either the inner or the outer boundary causes
\(\gamma\) to be slightly smaller (shallower), though in every case significantly different
from \(1/2\).

\subsubsection{Resolution study}
\begin{figure*}
  \begin{center}
    \resizebox{8 in}{!}{\includegraphics{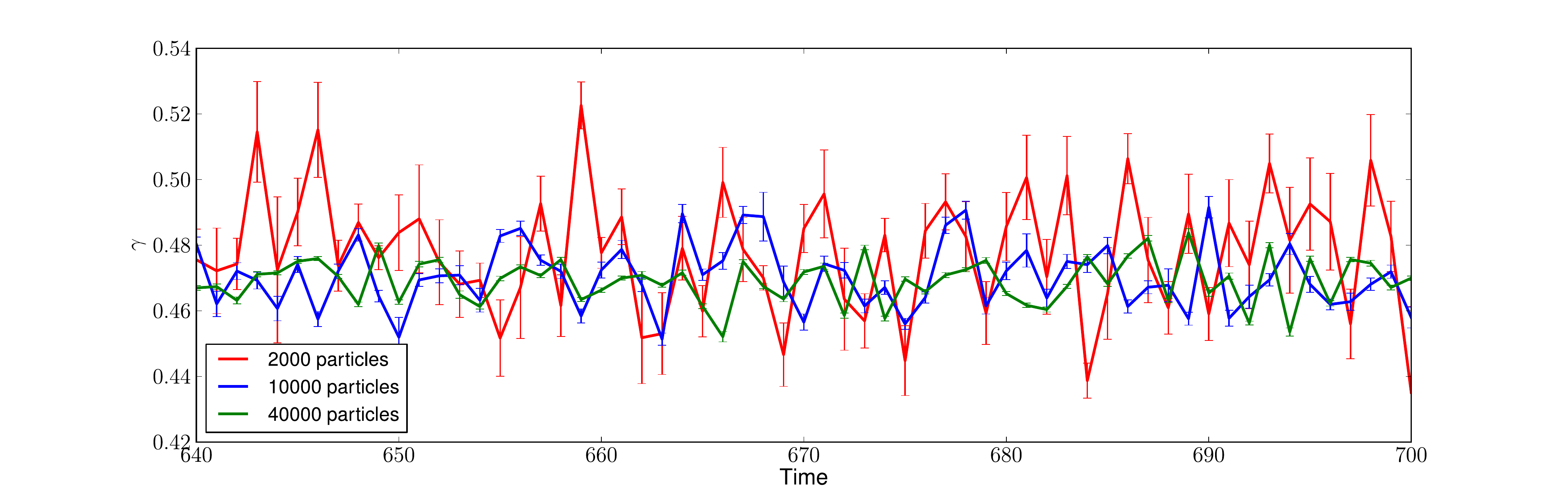}}
  \end{center}
  \caption{The best fit value of $\gamma$, with errors from the bootstrap procedure described in 
  section \ref{sec:fidsim}.  The variability from one snapshot to the next is higher than the bootstrap
  errors would suggest.  This is because the bin locations in the density estimator shift between 
  snapshots, but do not shift in the bootstrap fit to a single snapshot.  The time average value of 
  $\gamma$, and the rms fluctuation with time can be found in column 5 of Table \ref{tab:restable}.}
\label{fig:restime}
\end{figure*}

\begin{table*}
\centering
\begin{tabular}{|c|c|c|c|c|c|c|c|}
\hline
$N$ & $A$ & $\gamma$ & $\chi^2_{\rm red}$ & $<\gamma>_t$ & $\gamma_{\rm m}$ binned & $\gamma_{\rm m} $ single run & $\gamma$ from ${J(\mu)}$  \\
         2000 &        0.3499 $\pm$       0.0302 &        0.4345 $\pm$     0.0255 &       0.6307 &      0.4778 $\pm$       0.0187 &        0.4699 $\pm$     0.0009 &        0.4828 $\pm$     0.0003 &       0.4155 $\pm$     0.0008 \\
         4000 &        0.3048 $\pm$      0.0065 &        0.4602 $\pm$      0.0043 &       0.2583 &       0.4710 $\pm$      0.0152 &        0.4602 $\pm$     0.0007 &        0.4499 $\pm$     8e-05 &        0.4129 $\pm$     0.0004 \\
        10000&        0.3088 $\pm$      0.0043 &        0.4579 $\pm$      0.0031 &       0.3665 &      0.4703 $\pm$       0.0099 &        0.4567 $\pm$     0.0002 &        0.4378 $\pm$     1e-05 &        0.4115 $\pm$     0.0002 \\
        20000&        0.2865 $\pm$      0.0015 &         0.4720 $\pm$     0.0014 &       0.3393 &      0.4722 $\pm$       0.0089 &        0.4497 $\pm$     0.0001 &        0.4478 $\pm$     8e-06 &        0.4110 $\pm$       8e-05 \\
        40000&        0.2905 $\pm$     0.0007 &          0.4700 $\pm$     0.0006 &       0.3102 &       0.4691 $\pm$      0.0068 &        0.4491 $\pm$       7e-05 &        0.4606 $\pm$     4e-06 &       0.4107 $\pm$     4e-05 \\
\hline

\hline
\end{tabular}
\caption{We examine the sensitivity of the fits to the number of particles being used in the simulation.  
Each line represents six simulations with slightly different particle number, with our fiducial set of initial conditions.
Column 1 shows the approximate number of particles.    
Columns 2 and 3 are the best fit normalization and scaling for the 
functional form $\rho(x)=A\, |x|^{-\gamma}$.  
These were determined by fitting the mean of the densities of 6 separate runs,
over the range $10^{-3}<x<10^{-1}$, and 
bootstraping 10000 synthetic data sets to obtain the error bars.  Note that there is some significant covariance between $A$
and $\gamma$.  Column 4 shows the reduced $\chi_{\rm red}^2$.   
Column 5 shows the time average and variance of $\gamma$ for all time 
steps between $640<t<700$ (see also Fig. \ref{fig:restime}).  
Columns 6 and 7 show the inferred value of $\gamma$ obtained by fitting
the cumulative mass
(estimated from the data in two ways) over the same range in $x$, and Column 8 is inferred from a fit to the $J(\mu)$ relation in equation \refeq{eqn:Jvsmu}.  All 
fits were performed at the end of our simulation, at time 700, except the $J(\mu)$ fit, which has to be done at much earlier times 
because the relation disintegrates (see section \ref{sec:energy}).}
\label{tab:restable}
\end{table*}

In Table~\ref{tab:restable}, we show how these results depend on particle number
\(N\) over the range $2 \times10^3$ to $4 \times 10^4$.
Columns 2 and 3 show the best fit values of the  normalization \(A\) and exponent
\(\gamma\) of the density profile at time $t=700$,  Column 4 shows the 
reduced $\chi_{\rm red}^2$; this should be roughly unity if the data are consistent with the fitted
density profile and the errors are Gaussian.   However,  we expect $\chi_{\rm red}^2$ 
to be less than 1 because the errors in the 
density bins show significant negative kurtosis.  The apparent trend in $\gamma$ and $A$ with 
$N$ is illusory; it is only the case for this particular snapshot.  Although the smallness of the bootstrap errors 
indicate that the fit is very 
stable at a given snapshot in time, the value of the best fit $\gamma$ varies from one snapshot
to the next.   

Fig. \ref{fig:restime} plots the best fit value of $\gamma$ for several snapshots with errors from the 
bootstrapped data,  and demonstrates the extent of the time variability of the fit.  Simulations with more particles 
display smaller absolute rms variability with time, but have a much larger disparity between the 
rms variability in time and the bootstrapped errors in a particular time step.  This is because we have used a 
density estimator with approximately $20$ particles per bin for all resolutions in this study.  Therefore the simulations with more 
particles have more data points available for the bootstrap.  For all resolutions, however, the bin locations
shift from one time step to the next, keeping the number of particles per bin roughly equal to $20$.  This 
procedure generates variability in the density estimator from one snapshot to the next.   The time average
and variance of $\gamma$ over the interval $640<t<700$ are shown in column 5 of  Table \ref{tab:restable}.

As discussed in section \ref{sec:odg}, a self-similar solution exhibits its power-law behavior
in several different functional relations. In principle, the exponent \(\gamma\)
can be extracted from measurements of any one of these relations.
For example, the cumulative mass inside a position \(x\) is
$m(x)\propto |x|^{1-\gamma}$.
Since the mass is a cumulative variable, mass data points are
correlated; however we have opted to give all points equal weight in the 
bootstrap procedure. We have computed the cumulative mass in two ways, first by
numerically integrating the mean density (abscissae at the bin locations), 
and second by avoiding the binning entirely and tallying the total mass at
the particle locations for one of the runs (abscissae at the particle locations). 
The results of fitting the
cumulative mass can be found in columns 6 and 7 of Table \ref{tab:restable}. The exponents
are not consistent with the fits to the density, nor with one another. 
This discrepancy is not understood,
though it is easy to identify systematics which may affect the fit.
Being an integrated quantity,
the cumulative mass \(m(x)\) is sensitive to all regions of the data
interior to \(x\); in particular, it is not possible to excise the innermost part of
the object as we have done with the density.
Including more of the central region 
in the density fits will cause the power-law exponent 
to be systematically shallower than when it is excised, consistent with 
the mass result.
The binning also likely affects the fit differently for the mass and the density.

The most interesting of the indirect scaling relations is that expressed in
equation (\ref{eqn:Jvsmu}). The action of a particle orbit in the stationary mean-field
potential can be computed directly by integration over the orbit.  
The values for exponents obtained by fitting this relation are shown in column 8
of Table  \ref{tab:restable} and show a remarkable stability as a function of \(N\),
compared to the other methods.
In Fig.  \ref{fig:rhof1}, the density profile
corresponding to this value of the exponent is indicated by the dashed
line. The discrepancy between the direct fit to the density and the
indirect fit to the \(J(\mu)\) relation is very clear in that figure.
The details of the measurement and fit of the \(J(\mu)\) relation are
discussed further in section \ref{sec:energy}.

\subsubsection{Sensitivity to initial conditions}
\begin{figure}
  \begin{center}
    \resizebox{3.6 in}{!}{\includegraphics{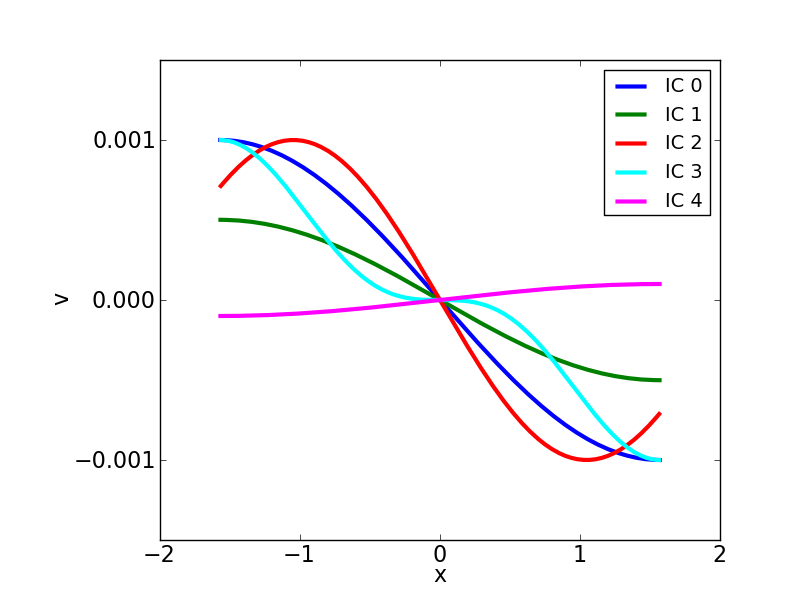}}
  \end{center}
  \caption{Four variations in the initial velocity perturbation. The
    density profiles arising from these initial states are shown in
    Fig.~\ref{fig:icf}, and an intermediate snapshot of the phase space
    configuration is shown in Fig.  \ref{fig:imsnap}.  The functional forms are:
   [IC 0] $\upsilon_i(x_i)=-0.001 \sin(x_i)$, [IC 1] $\upsilon_i(x_i)=-0.0005 \sin(x_i)$, 
   [IC 2] $\upsilon_i(x_i)=-0.001 \sin(1.5*x_i)$,
    [IC 3] $\upsilon_i(x_i)=-0.001 \sin^3(x_i)$ and [IC 4] $\upsilon_i(x_i)=+0.0001 \sin(x_i)$.
    \label{fig:ici}
  }
\end{figure}

To close this section on simulations of cold collapse, we discuss the
sensitivity of the exponent \(\gamma\) to changes in the initial conditions (ICs).
Fig.~\ref{fig:ici} shows
several of the initial conditions that we have tested. These are all cold and
homogeneous in density, but with a variety of initial velocity
perturbations.
IC 0 is the fiducial case, with particle velocities given by equation
(\ref{eqn:initvel}) with $V_0=0.001$. IC 1 tests the sensitivity to the
velocity amplitude; IC 2 investigates the impact of a
slow group of outer particles whose infall velocity lags 
that of intermediate particles; IC 3 has two additional inflection points away from the origin;
IC 4 examines initially expanding rather than initially contracting particles.

\begin{figure}
  \begin{center}
    \resizebox{3.6 in}{!}{\includegraphics{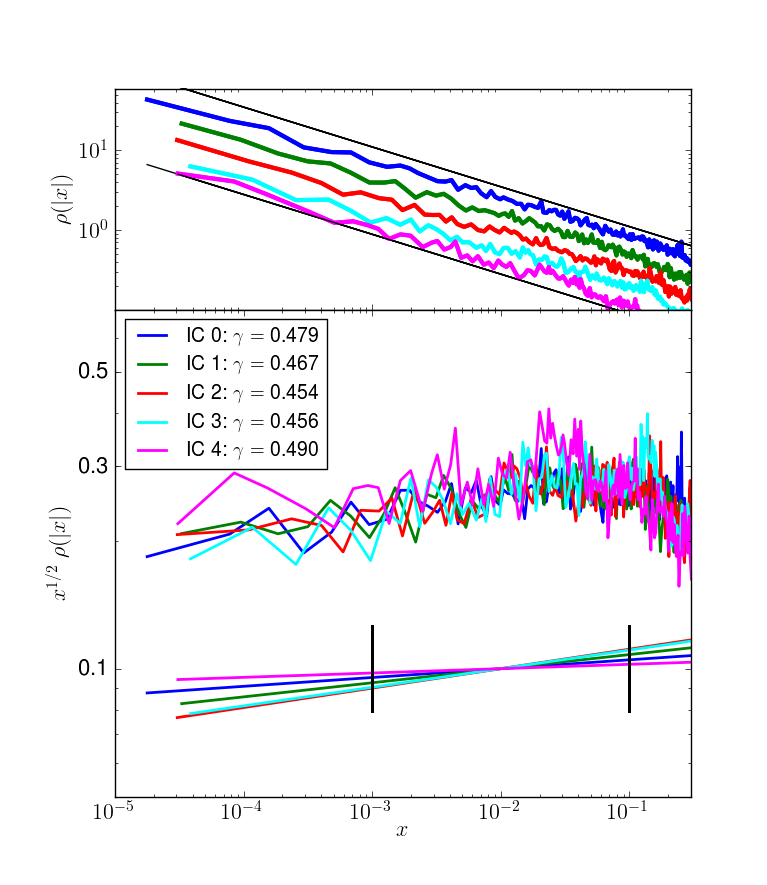}}
  \end{center}
  \caption{ {\bf Top:}  The density profiles at $t=700$ for the initial conditions plotted in Fig.  \ref{fig:ici}.
  These curves are averages over 6 simulations of $\sim 10000$ particles, 
  with slightly different particle numbers. The curves have been shifted
  vertically for visual clarity.  The two solid black lines show $x^{-1/2}$ power laws for comparison.  {\bf Bottom:}
  The same densities scaled by $x^{1/2}$, and not artificially shifted. This panel demonstrates more clearly that
  the attractor solution is somewhat shallower that $\rho \propto x^{-1/2}$. The short vertical lines
  indicate the range of data that were used to fit the exponent and amplitude of the density.  The lower curves
  show the best fit power law for each initial condition, also indicated in the legend. This helps quantify the
  extent to which the density approaches an attractor.  We caution the reader that there is some covariance
  between the slope and the amplitude in the fit.}
  \label{fig:icf}
\end{figure}

Fig.  \ref{fig:icf} shows the final configuration ($t=700$) of each IC.  The top panel
shows the average density of six simulations with the same number of particles as for 
the fiducial simulation,
for each initial condition.
The  normalizations of all ICs except IC 0 have been artificially shifted for visual clarity.
Solid black curves with power-law behavior \(x^{-1/2}\) are also shown for reference.   
The bottom panel shows the same average densities, but with an \(x^{-1/2}\) trend
scaled out. The normalizations are not shifted in this panel. We see that the resulting 
density profiles for these different initial conditions are consistent
with an attractor behavior \(\rho(x) \propto |x|^{\gcrit}\) with \(\gcrit\simeq0.47\).
Power laws with the best fit $\gamma$ (in the legend) are plotted in the lower part
of the bottom panel, to help quantify the extent to which the density approaches an attractor.
The vertical black hashes show the range of data that was used in the fit. There is some
covariance between the slope and the amplitude.

\begin{figure}
  \begin{center}
    \resizebox{3.6 in}{!}{\includegraphics{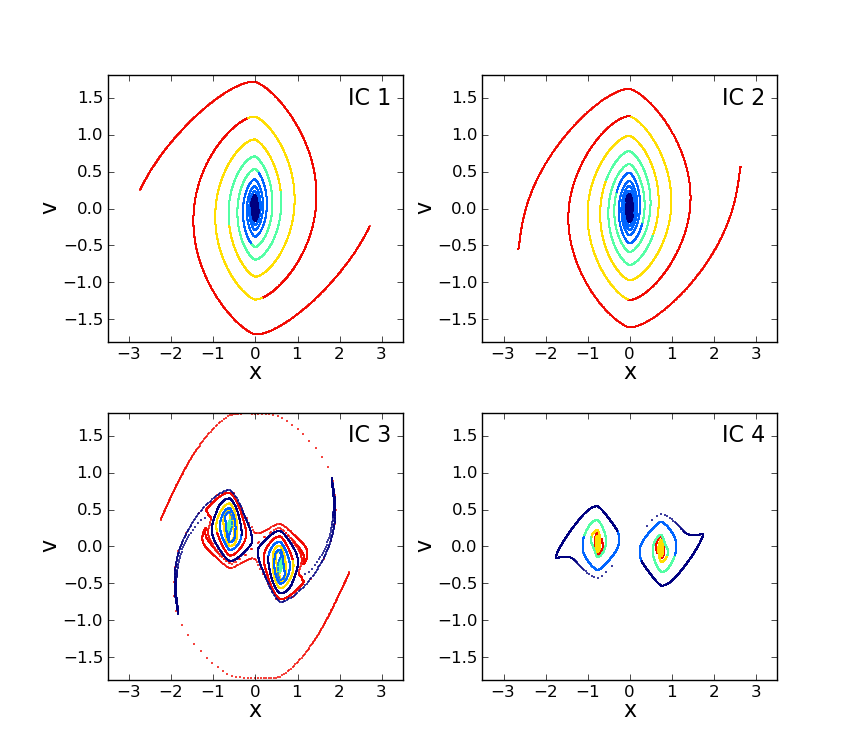}}
  \end{center}
  \caption{An intermediate snapshot at time $t=40$ in the evolution of ICs 1-4.  ICs 1 and 2
    undergo monolithic collapse while ICs 3 and 4 first form two distinct
    objects that later merge to form the final virialized object.  The color
    coding indicates the initial positions of particles.  Blue particles
    started at small distance and red started at large distance.
  }
  \label{fig:imsnap}
\end{figure}

Even though the final states of these simulations are similar, there are large qualitative differences in
the formation history. Fig.  \ref{fig:imsnap} shows an intermediate snapshot
of the phase-space configuration for these four simulations; ICs 1 and 2
undergo a monolithic collapse, while ICs 3 and 4 each form two distinct halos
that later merge to form the final product.  These can be compared with the bottom left panel of 
Fig.  \ref{fig:evsimple}, which shows IC 0 at the same time.  The colors in Fig. 
\ref{fig:imsnap} encode the initial positions of the particles; blue particles
were initially located close to \(x=0\) and red were at the largest distances.
Note that the inner particles remain in the interior of the final halo only
for cases that undergo monolithic collapse. 

\subsection{Warm initial conditions}
\label{sec:density:warm}
The self-similar final states discussed above arise from cold initial conditions. At the 
end of Section \ref{sec:odg}, we
described ``coldness" as localization of the particle density to a single 
smooth curve in the \((x,\upsilon)\)-plane.
In this section we describe simulations whose final states are quite different
from the self-similar states described above.
We assume in this discussion that a state initialized with velocity perturbations
from a stochastic process (noise) satisfies any conceivable definition of ``warm''.
Our interest is in the approach to warmness as structure is added to the function
describing the cold initial particle locus.  We shall find that 
even a small amount of additional structure in the initial state, 
that appears consistent with the notion of ``coldness," destroys the self-similarity
of the final state.  

In a particle representation, the degree of smoothness of
the particle locus must be compared to the particle separation. In the
cold scenarios above, initial velocity perturbations were chosen to
have wavelengths much larger than the initial particle separation.
As the wavelength of the initial velocity perturbation decreases,
it approaches the Nyquist limit for particle sampling, becoming
more like a velocity noise in character.
In this limit, we expect the subsequent evolution to resemble that of
warm initial conditions, where the velocity perturbations are generated
by independent stochastic processes for each particle. 
As a consequence, we expect that the resulting equilibrium halo
will develop a core; collapse from a state of nonzero velocity dispersion
should develop a core because the maximum phase-space density cannot increase
\citep{1979PhRvL..42..407T}.

\begin{figure}
  \begin{center}
    \resizebox{3.6 in}{!}{\includegraphics{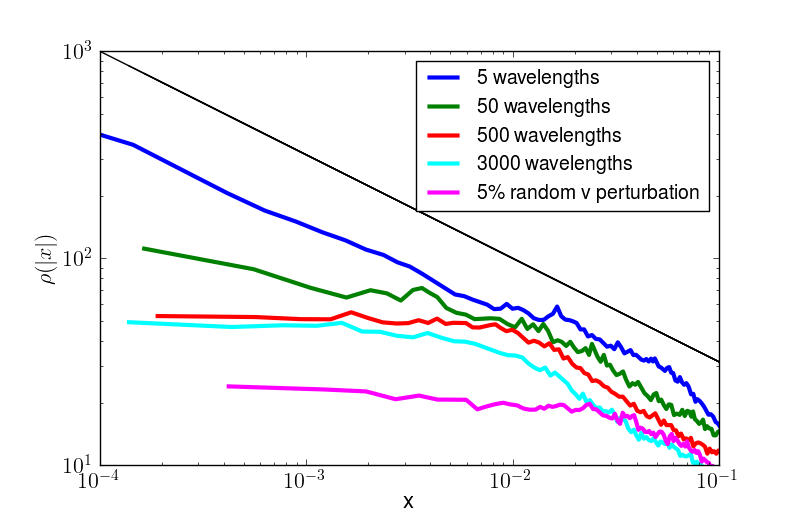}}
  \end{center}
  \caption{
    A short wavelength component in the velocity perturbation makes
    the final density profile depart from a power law behavior.  The caption indicates 
    the number of wavelengths over the interval $(-\pi/2,\pi/2)$ in the initial conditions.  
    As the wavelength approaches the Nyquist wavelength, equal to twice the initial
    inter-particle spacing, the simulation behaves more and more like a
    warm simulation (magenta curve) in which the initial velocities have been perturbed with
    5\% random noise.  These curves are averages over 6 simulations, and have been further
    averaged over 20 time units. The time averaging is used to smooth the impact of substructure
    on the density profile. The short wavelength component generates long lived substructure 
    (cf. Fig. \ref{fig:subst})
    that does not disappear over the timescales that we ran the simulations.}
  \label{fig:warm}
\end{figure}

Fig.  \ref{fig:warm} shows the density profiles
that result from adding an extra component to the velocity perturbation of
equation~(\ref{eqn:initvel}).
The additional component has a lower amplitude than the original perturbation (\(0.1 V_0\)), but a higher
wavenumber modulation. We have tested components containing 5, 50, 500 and 3000 wavelengths over the
domain of the data; for comparison, the Nyquist rate (determined by the initial inter-particle spacing)
would correspond to $N/2=5000$ wavelengths over the domain  of the initial particle
distribution ($-\pi/2 \le x \le \pi/2$).
The bottom curve (magenta) shows a warm simulation in which an
RMS 5\% random fluctuation has been added to the initial velocities.  These
curves are averaged over six simulations and 20 time units (in the range \(t=\) 680-700) for
reasons explained below. As expected, the density profile becomes
shallower near the center as the spatial frequency of the initial velocity perturbation increases.
These features indicate that it is difficult to achieve coldness in particle realizations of the 
continuum.  It is remarkable that a well-defined constant-density core appears even in a 
simulation for which the spatial frequency is only 10\% of the Nyquist frequency.

\begin{figure*}
  \begin{center}
    \resizebox{7 in}{!}{\includegraphics{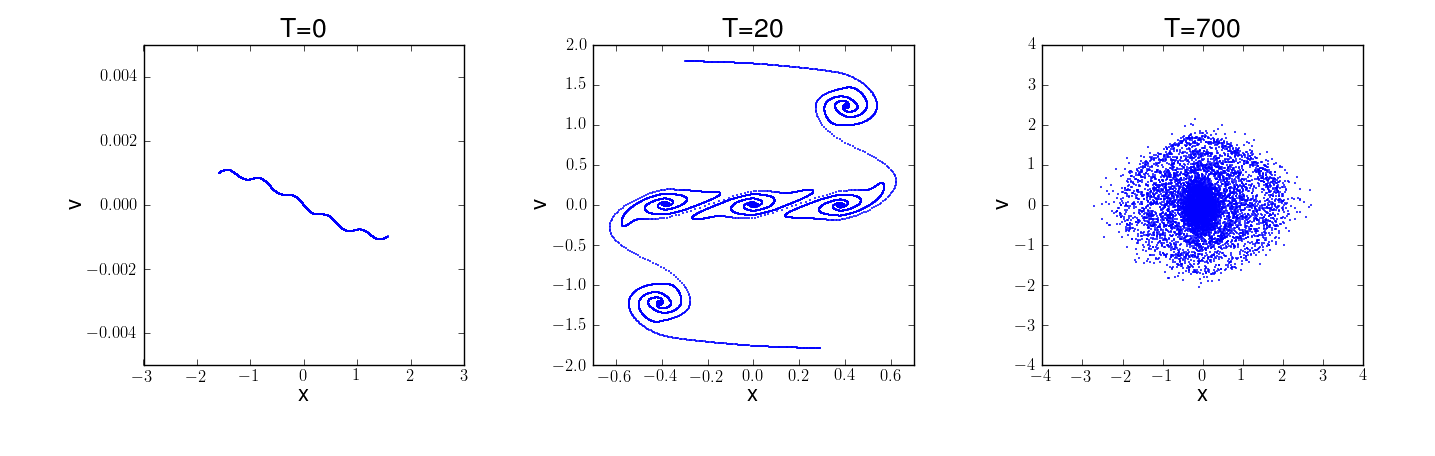}}
  \end{center}
  \caption{Three snapshots from the simulation leading to the top (blue) 
    density profile in Fig.  \ref{fig:warm}.
    The small velocity perturbation results in the formation of several
    distinct objects that later merge to form the final product.
    The average central density at late times is less than that 
    in simulations without small scale power.}
  \label{fig:subst}
\end{figure*}

Less obviously,
a high frequency perturbation in the initial velocities leads to the formation of substructures.
This is illustrated in Fig.  \ref{fig:subst}, which shows three successive
times from the simulation plotted in blue of Fig.  \ref{fig:warm}, in which the initial 
velocity perturbation contains five wavelengths in the domain of the initial particle distribution.
Many individual sub-halos form in locations where the flow is initially 
converging, before merging to form the final
product. The sub-halos are very long lived; some of them never dissolve 
over the simulation times we have tested.  Thus we can only measure a density
profile in a time averaged sense---this is why 20 time units
have been averaged to produce the curves in Fig. ~\ref{fig:warm}.

\subsection{Cold initial conditions with inhomogeneous density}
\label{sec:density:steep}

All simulations discussed thus far were initialized with constant density.
In this section we report on simulations initialized with an inhomogeneous
density. In particular, we choose an initial power-law density profile
\(\rho_{\rm init}(x) \propto|x|^{-\gamma_i}\). We will demonstrate
an apparent bifurcation in the form of the final state as a function
of the initial exponent \(\gamma_i\).
The simulations described here all had \(N=10006,\,10012,\,10018,\,10024,\,10030\) and \(10036\)
particles, with particle spacing chosen so that the resulting density is
a power law with exponent \(\gamma_i\).

\begin{figure}
  \begin{center}
    \resizebox{3.6 in}{!}{\includegraphics{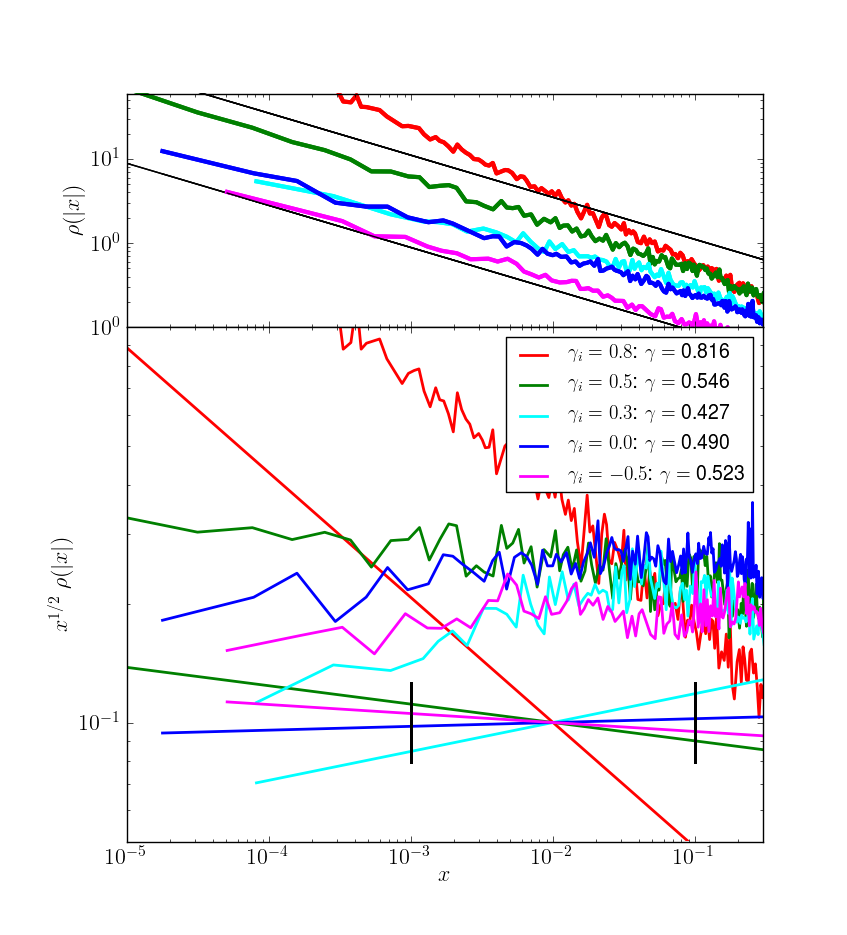}}
  \end{center}
  \caption{{\bf Top:}  The final density profiles for initial conditions that follow a
    power law $\rho \propto |x|^{-\gamma_i}$.  The final power law $\gamma$ is also noted 
    in the caption.  The curves have been artificially shifted for visual clarity.  
    The solid black lines show $|x|^{-1/2}$ power law for reference.
    {\bf Bottom:} The same densities scaled by $x^{1/2}$, and not artificially shifted.
    Initial power-law profiles that are shallower than $\rho(x)
    \propto |x|^{-1/2}$ evolve toward an approximate attractor, but initial profiles that
    are steeper, such as the red curve, retain their initial power law. }
  \label{fig:plf}
\end{figure}

Fig.  \ref{fig:plf} summarizes the results of a set of simulations with
initial exponents \(\gamma_i = 0.8, 0.5, 0.3, -0.5\).   The curve labeled $\gamma_i=0.0$
shows our fiducial simulation for comparison.  All initial particle
velocities were set to \(\upsilon=0\) in these simulations (except the fiducial case).
We find that initial conditions with slope \(\gamma_i\lesssim \gcrit\) collapse to form
final states with roughly the same \(|x|^{-\gcrit}\) density profile as homogeneous
initial states, shown in dark blue in Fig.  \ref{fig:plf}. However, initial conditions with
\(\gamma_i\gtrsim \gcrit\), such as the red curve in Fig.  \ref{fig:plf}, all preserve their density 
profiles.  The bottom panel of Fig.  \ref{fig:plf} shows that the attractor 
is not as tight as in the homogeneous case; the curves do not overlay one another as tightly as
in Fig.  \ref{fig:icf}, and there is more scatter in the measured power law index.  
Also the central portions of the 
profile appear to be flatter than the behavior at larger \(|x|\).  For this reason
the fit values (displayed in the legend) are somewhat more sensitive than fits in Fig.  \ref{fig:icf}  
to the range of data used
in the fit (indicated in the bottom panel of Fig.  \ref{fig:plf} with short vertical black lines).

\begin{figure}
  \begin{center}
    \resizebox{3.6 in}{!}{\includegraphics{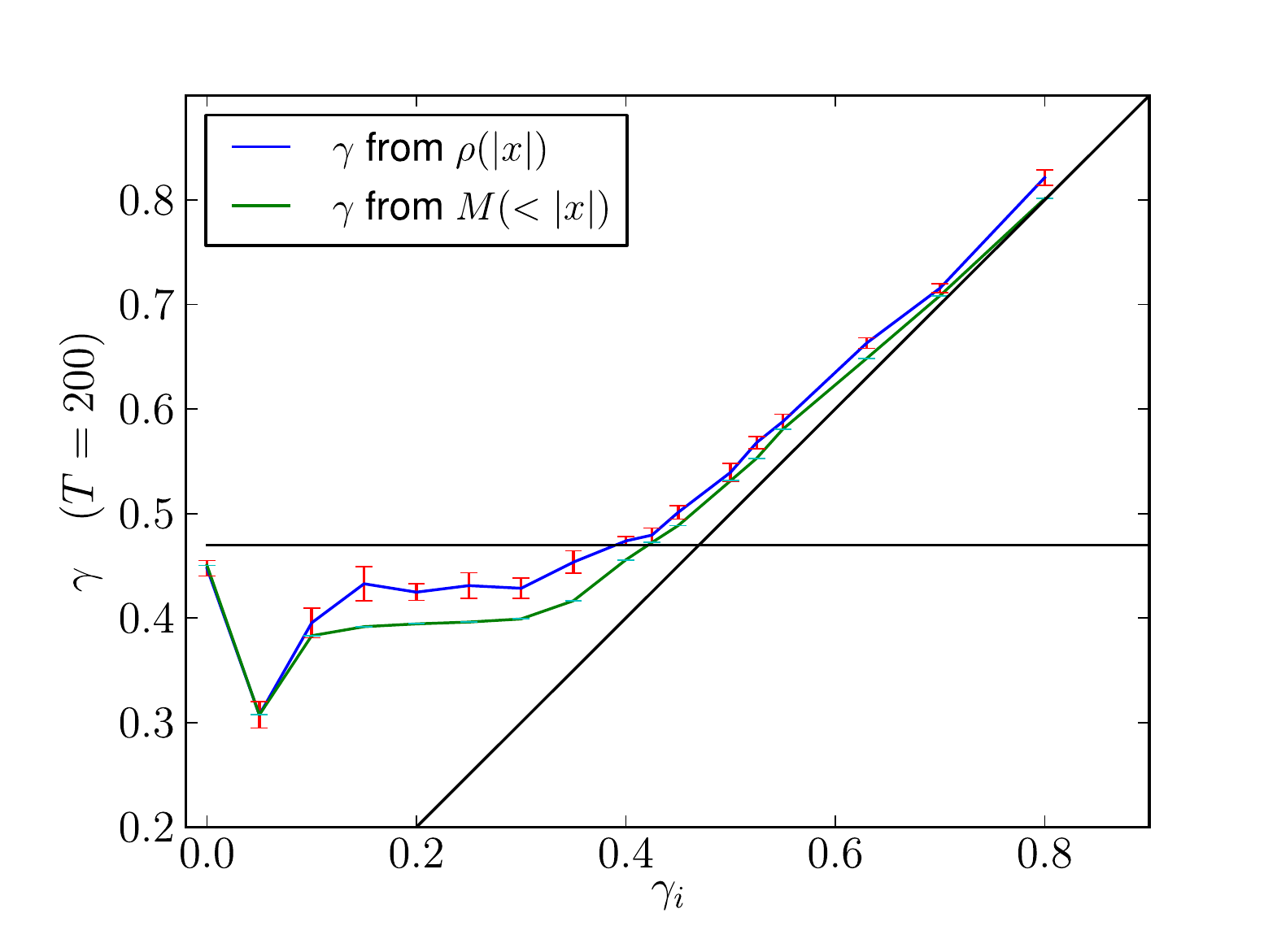}}
  \end{center}
  \caption{The final exponent $\gamma$ of the density profile $\rho(x) \propto |x|^{-\gamma}$
  at time $t=200$, 
  as a function of the initial exponent $\gamma_i$ in the initial inhomogeneous distribution of 
  particles $\rho_{\rm init}(x)\propto |x|^{-\gamma_i}$.
  Also shown in green are fits using the integrated mass interior to $|x|$. 
    Initial slopes steeper than the attractor solution $\gamma_i > \gcrit$ tend to steepen only
    slightly. In contrast, initial slopes much shallower than the attractor
    $\gamma_i< \gcrit$ steepen considerably.}
  \label{fig:scaling}
\end{figure}

Fig.  \ref{fig:scaling} shows the exponent of the final density profile, $\gamma$,
versus that of the initial condition for a larger set of values of \(\gamma_i\).  Each data point is 
computed using 6 simulations. 
For these simulations, we have perturbed the particle velocities according to equation \refeq{eqn:initvel},
because this significantly decreases the run time. For the cases with $\gamma_i=0.8, 0.5, 0.3, -0.5$,
we have verified that the velocity perturbation does not significantly impact the measured slopes.
Solid lines for \(\gamma=\gamma_i\) and \(\gamma=\gcrit\)
are also plotted for reference.

Although the bootstrap method significantly reduces the sensitivity to the
range of data used for the fits in Fig.  \ref{fig:scaling}, it does not eliminate it entirely.
If either the inner or outer boundary of the domain
is reduced by an order of magnitude or more, the best fit \(\gamma\) becomes systematically 
shallower. The fitted exponents in Fig.  \ref{fig:scaling} are
power-law fits to the portion of the density in the range $10^{-3} < |x| < 10^{-1}$.
The figure suggests that a system initialized in a more collapsed state than the attractor
solution remains close to its initial power law.  However,
a system initialized with a shallower density profile than the attractor significantly steepens
to evolve toward the attractor.
The dip of very shallow slopes for \(0 < \gamma_i < 0.2\) has its origin 
in two features of these final states.  First, at \(t=200\) 
the simulations with \(0 < \gamma_i < 0.2\) display noticeable gaps in phase space structure
in the range \(10^{-3} < |x| < 10^{-1}\).
These gaps, and similar gaps in the homogeneous simulations, are discussed below.
These gaps affect the fit for the exponent. Second, 
simulations with \(0 < \gamma_i < 0.2\) achieve much lower peak phase-space densities
than the homogeneous case or cases with \(\gamma_i > 0.2\).  We postulate that either the
gaps or other discreteness effects interfere with the efficient transfer of energy from the inner
particles to the outer ones.  
Since the inner threshold is fixed, we are not discarding the very innermost portion of the 
data in these simulations, and the fits result in systematically
shallower values of \(\gamma\), as discussed above.

\subsection{Time evolution of rank, action, energy and density}
\label{sec:energy}

\begin{figure*}
  \begin{center}
  \includegraphics[trim=2cm 0cm 2.0cm 0cm, totalheight=6.2 in]{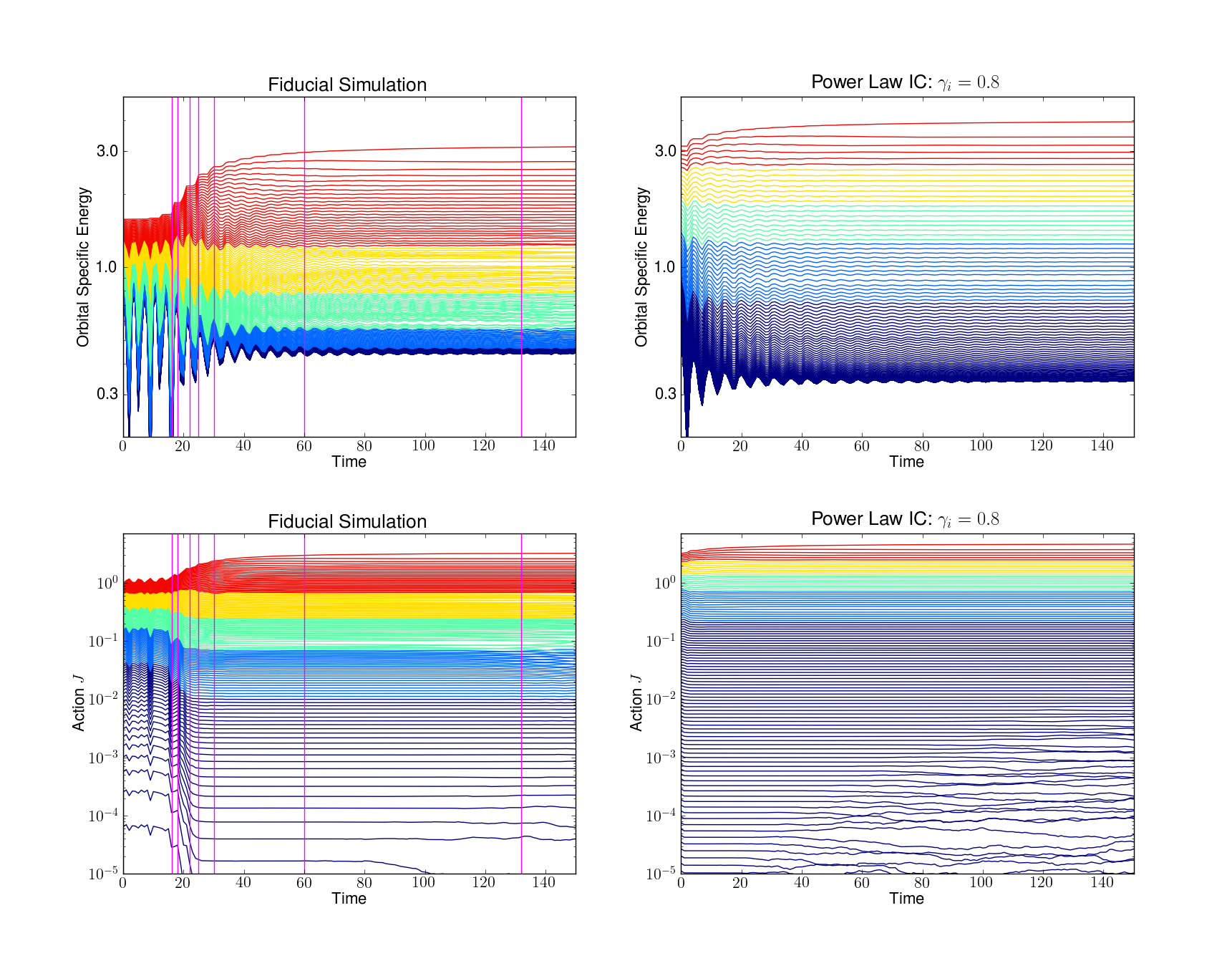}
  \end{center}
  \caption{A comparison between the fiducial simulation, initialized with a homogeneous 
  density, and a simulation in which the initial configuration of particles follows a steep
  power law density profile.  {\bf Top row:}  The evolution of the energy of every 40th particle in the
     fiducial simulation, and the simulation with a steep initial power law configuration
     with $\gamma_i=0.8$.
   The period of violent relaxation in the fiducial simulation can be
    seen between times 0 and $\sim$50.  After this time the gravitational potential
    is approximately stationary and the particle energies hardly evolve.  The
    vertical lines indicate times shown in Fig.  \ref{fig:evol}.
    {\bf Bottom row:}  The actions for the fiducial and power law simulations, 
    derived by assuming a stationary potential and 
    performing the integral of equation \refeq{eqn:actiondefinition}.
    The fluctuations for low $J$ at late times are almost certainly caused
    by discreteness noise.
  }
  \label{fig:evst}
\end{figure*}

\begin{figure*}
  \begin{center}
    \includegraphics[trim=2cm 0cm 3.0cm 0cm, totalheight=0.7\textheight, angle=90]{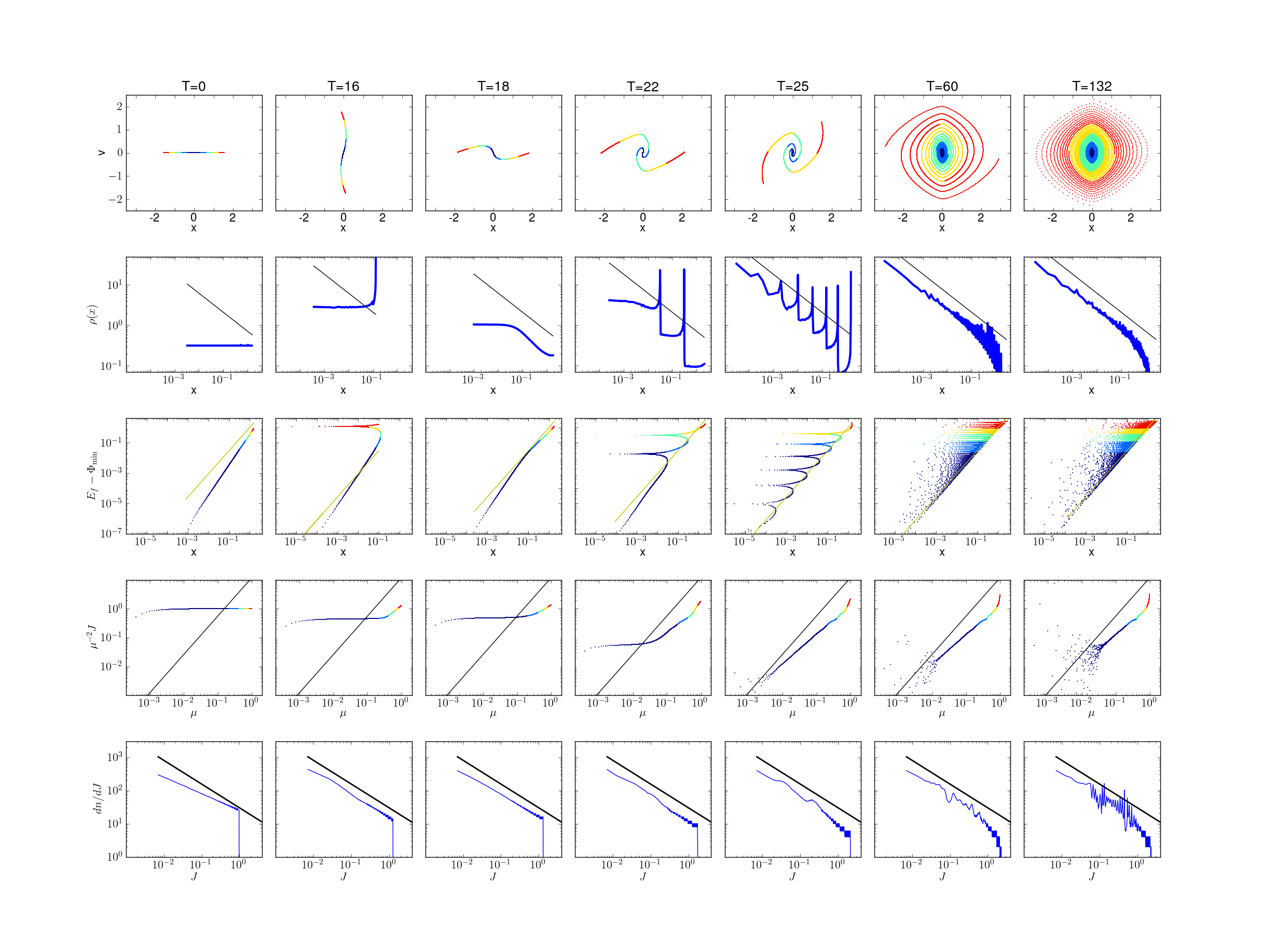}
  \end{center}
  \caption{
    Eight snapshots of the fiducial simulation are shown. The time
    is given at the top of each column; each time is indicated in Fig. 
    \ref{fig:evst} by a vertical line. The top row shows the phase-space
    distribution of the particles.  The colors indicate 5 quintiles in initial
    particle position, blue started near $x=0$ and red started far away.
    The second row shows the density profile.  The diagonal black line is
    $x^{-1/2}$. The third row shows
    a scatter plot of the energy and current
    position, the colors again indicating quintiles in initial position.
    For density $\rho \propto |x|^{-\gcrit}$, the potential varies as 
    $|x|^{2-\gcrit}$ and this is shown with a solid green line using $\gcrit=0.47$.
    The fourth row shows the measured \(J(\mu)\) relation (equation \ref{eqn:Jvsmu}).
    The bottom row shows the distribution of the number of particles per unit 
    \(J\).}
\label{fig:evol}
\end{figure*}

Fig.  \ref{fig:evst} shows the evolution of the particle energies and actions as a
function of time. The different colors mark quintiles in the initial particle
position, as described in Section \ref{sec:sim}.  
The evolution of the energies for the fiducial simulation, in the top left corner,
shows rapid oscillations for $t\lesssim 50$.  The phase space plots of the system 
configuration can be seen in Fig. \ref{fig:evol}, at the times indicated by the vertical
magenta lines.  Before time $22$, the system 
is undergoing rod-like rotation in phase space, 
with only one stream per location $x$.  The oscillations 
in the energy are dominated by changes in the potential, as the system oscillates between 
maximum rarefaction and maximum compression.  At 
approximately $t = 22$, the system transitions from one to three streams at small $x$.
After this time the central part of the particle locus winds up quickly, with more than $15$ streams 
in the inner region by time $t=25$, but the oscillation pattern 
in energy continues in phase with the outer particles, suggesting that the potential in the center
is dominated by the configuration of the particles in the outskirts at these times.
The range of particle energies spreads with time, indicating
that energy is being transferred from inner to outer particles. The colors are
not mixed, demonstrating that the particles approximately maintain their rank
ordering in energy; violent relaxation redistributes energy among the
particles but does not scramble the energy ordering.  For \(t\gtrsim 50\) the
energies of the particles are nearly constant.  For
example, an extended run shows that the energies of the (most, least)
energetic particles change by (0.5\%,1.0\%) between 150 and 700 time units.
This simulation was run for several values of
\(N\), as indicated in Table \ref{tab:restable}. The energy oscillations from
violent relaxation occur in phase for all of these runs, and the final
energies of the (innermost, outermost) particles of the two highest resolution
runs match to within  (0.7\%, 0.6\%) in all the simulations.

The time evolution of the particle actions in the fiducial simulation is shown
in the bottom left hand panel of Fig.  \ref{fig:evst}.  The actions are computed
assuming the potential is fixed at its instantaneous functional form, even though for times
less than $t=50$ the potential is evolving significantly.  The fluctuations in low $J$ 
at late times are almost certainly caused by discreteness noise.  The actions display a number of 
interesting features.  First, they stabilize extremely abruptly, 
well before the energy oscillations damp out and the energies of the particles cease
to evolve.  Second, the transition to constant action for the inner particles 
happens just after the transition from rod-like 
rotation to multiple stream flow.  This can be seen in the fiducial simulation by comparing 
to the top row of Fig. \ref{fig:evol}, which shows that the transition from rod-like rotation 
to multi-stream flow happens at approximately time $t=22$.

Comparing the energy and action plots we learn that 
the relaxation of the system appears to be a two stage process.  The first stage, in
which both the energies and the actions evolve rapidly, happens primarily before the transition from 
one to three streams.  The second stage, in which the actions rapidly settle and the energies 
evolve adiabatically, happens as the system winds up, primarily between the one-to-three 
stream transition, and the ultimate stabilization of the outer gravitational potential.  During this second
stage, the amount of energy transferred between particles, and the distances over which it can be 
transferred both tend to zero as the number of streams increases and the inter-caustic distance decreases.

\begin{figure}
  \begin{center}
    \resizebox{3.6 in}{!}{\includegraphics{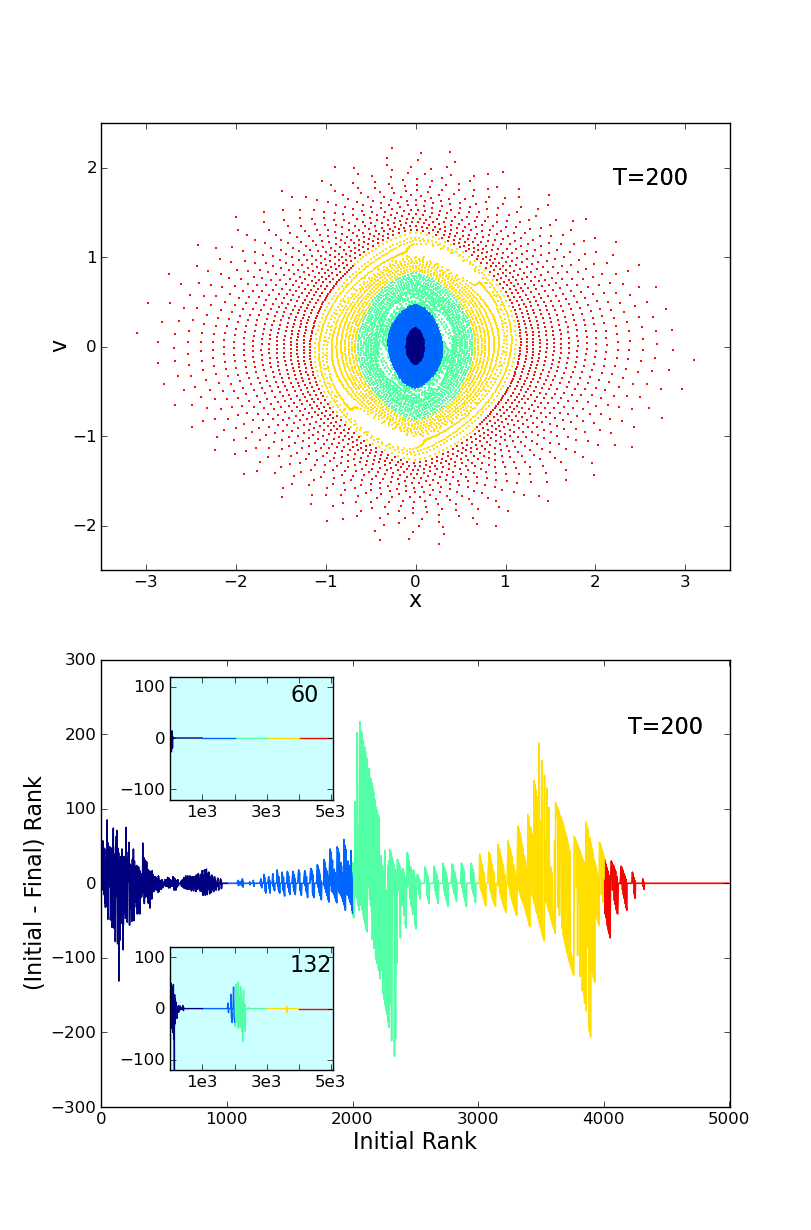}}
  \end{center}
  \caption{{\bf Top:} The gaps developing in the energy distribution in 
  Fig.  \ref{fig:evst} 
    correspond to empty pockets in phase space.  This panel shows  
    gaps in the yellow and green marked particles, at
    energies $E-\Phi_{\rm min}\approx 0.6$ and $E-\Phi_{\rm min}\approx 0.9$.
    {\bf Bottom:} The energy gaps
    cause the rank ordering of the particles to be scrambled, in certain
    bands.  The insets show the rank ordering at two earlier times, also
    studied in Fig.  \ref{fig:evol}.
   }\label{fig:gappy}
\end{figure}

The panels on the right of Fig.  \ref{fig:evst} show the evolution of the particle energies and actions
for one of the simulations initialized with a steep power law density profile.  These plots contrast starkly
with the evolution of the fiducial simulation.  We see that the actions (bottom right panel) are immediately stable, and evolve 
very little from their initial values.  This is the case because the one-three stream transition occurs well 
before time $t=1$, and the system does not undergo several rod-like rotations before the transition to multiple 
streams.  Therefore, steep systems like this one relax mainly through the second adiabatic stage of energy
redistribution.  The redistribution of particle energies (top right panel) 
is much less than in the fiducial case.  Since the particles are essentially in fixed stable orbits
from time $t=1$ onward, the change in the scaling of the density profile is small, as can be seen in Fig.  \ref{fig:scaling}.  

A curious feature in the top left panel of 
Fig.  \ref{fig:evst} is the development of narrow gaps
in the energy distribution, seen after about 60 time units, in the yellow and
green curves.
The gaps correspond to specific phase space pockets, empty of particles,
which open at times significantly after the end of violent relaxation.
An example can be seen in the top panel of Fig.  \ref{fig:gappy}.
The emergence of these gaps scrambles the
energy rank-ordering of the particles that had been preserved in the earlier
stages of the evolution (bottom panel).

These gaps are a robust result of our simulations.  
We observe them in simulations with
all values of \(N\), with and without imposing
explicit left/right symmetry, and gaps appear in both the heap and the
search-tree versions of the code run on identical initial
conditions.  They also occur in the simulations initialized with a power law
density profile, though they manifest at different times and are much less
pronounced when the initial density profile is steep.

Fig.  \ref{fig:evol} shows the state of the fiducial system at 8 specific
times, which are indicated in the left panels of Fig.  \ref{fig:evst} with vertical lines.
The phase-space distribution (first row) is shown with color indicating
quintiles in initial position. The evolution of the density profile (second row)
shows the steady growth in the number of caustics. At early times the density is roughly
constant in the intervals between caustics. As the evolution proceeds, the caustics
become more closely spaced and the usual \((x_{\rm max}-x)^{-1/2}\) density
distribution associated with a fold singularity at \(x_{\rm max}\) dominates the
behavior between caustics.  We also notice that the caustics are roughly
equally spaced in logarithmic distance.

Also shown in Fig.  \ref{fig:evol} are plots of energy versus position for the
particles (third row). The colors in this row again indicate quintiles in initial particle
position. As seen in these plots, violent relaxation mixes the final positions
relative to the initial ones (the colors are mixed when projected onto the
horizontal axis) but does not mix the final energies relative to the initial
ones (the colors are not mixed when projected onto the vertical axis).
The envelope of the energy-position plot gives the maximum
orbital excursion as a function of energy; the locus of this
envelope is the potential as a function of
distance, and therefore displays the power-law behavior \(|x|^{2-\gcrit}\).
For comparison, this form of the potential is shown with a green line.  
Indeed, if the turnaround is known for all the particles, this behavior can
be used as another method to infer \(\gamma\).  We examined this for the fiducial 
simulations and found that the power law is slightly shallower than when fitting the
cumulative mass, though not as shallow as when fitting \(J(\mu)\).  Since the 
potential is a double integral of the density, it displays correlations and
dependency on the global structure of the state, similar to the effects
discussed for the cumulative mass scaling in section \ref{sec:results:cold}.

The evolution of the \(J(\mu)\) relation is displayed in Fig.  \ref{fig:evol} (fourth row) and shows
a very clean power law developing even before the end of violent relaxation ($t=25$).  Here we have scaled 
by $\mu^{-2}$ so that the initial relation is flat.  The solid black line shows the scaling from
equation \refeq{eqn:Jvsmu} for the value $\gcrit=0.47$, demonstrating that the asymptotic scaling
of this relation predicts a different value of $\gamma$ than the other scaling relations. 
The power-law behavior develops much earlier than the time at which power-law behavior in the
density becomes apparent.  The panel at $t=22$ shows that this power-law behavior is 
first seen in the outer parts of the system and last established in the inner parts, 
which is perhaps somewhat counterintuitive. This
dynamical behavior is related to the fact that the phase-line remains close to a straight
line longest at \(x=0\). The clean power law in \(J(\mu)\) deteriorates at later times, apparently 
due to the loss of coherence in the center of the system, consistent with the loss of correlation
between $\mu$ and particle energy.

As already indicated in Table \ref{tab:restable}, the value of the exponent \(\gamma\)
extracted from the measured \(J(\mu)\) relation is systematically lower than that
extracted from fits to the density. However, these exponents cannot be compared
at the same times, because the \(J(\mu)\) relation deteriorates before the
density is sufficiently settled. All fits to the \(J(\mu)\) relation referred to in this
paper have been done at times around \(t=30\) to allow sufficient dynamic range
for an accurate fit, \(10^{-2} < \mu < 10^{-1}\). However, the observed slope of \(J(\mu)\)
does not appear to evolve at all past this time, and it is easy to conjecture that
a simulation with sufficient resolution in the center to delay the deterioration
would show no subsequent change or drift in the exponent determined from \(J(\mu)\).
We conjecture that the reason the exponent obtained from \(J(\mu)\) does not agree with
fits to the density is because the monotonicity of the relation is destroyed by discreteness 
effects (see bottom panel of Fig. \ref{fig:evst})
well before the system is sufficiently wound to render the expression in equation
\refeq{eqn:Jvsmu} valid. 

The bottom row in Fig.  \ref{fig:evol} shows the evolving distributions of orbital
actions (per unit $J$); we expect a coarse-grained power-law behavior as expressed in
equation \refeq{eqn:action}.  This is shown in the plot with a solid black line for $\gcrit=0.47$. 
The observed distribution shows clear systematic
deviations from the expected power law at intermediate times and becomes sufficiently disordered
at later times that it is impossible to extract a meaningful value of \(\gamma\).
Although we might expect certain subtleties in the convergence of the
distribution in \(J\), it is somewhat surprising that this distribution
can display such complex structure, while the observed \(J(\mu)\) relation
is quite stable and clean.

We do not show similar snapshots for the simulations initialized with a
power-law density profile; however, we now describe the results qualitatively.
The violent relaxation and settling to an equilibrium gravitational potential
proceed much faster in simulations with a steep initial profile compared to
those with a shallow initial profile. The development of gaps of the type shown in
Fig.  \ref{fig:gappy} is strongly suppressed in the steep cases, compared to the
shallow cases.
Finally, there is a difference in the evolution of the distribution of actions:
the shallow cases eventually reach an \(f(J)\) distribution that follows equation
\refeq{eqn:action} with respect to the final slopes of the density profiles.
However, the steep cases do not. The \(f(J)\) distribution for the steep cases
begins very close to the \(f(J)\) that would be expected for a \(\rho \propto
x^{-\gcrit}\) scaling, and since the actions retain their initial values (shown in Fig.  \ref{fig:evst}), 
this distribution does not evolve at all. 

\section{Conclusions}
\label{sec:conclusions}

We have performed a suite of simulations of one dimensional gravitational
collapse, equivalent to \(N\) infinite parallel mass sheets collapsing along
one axis in three dimensions. We find that a self-similar attractor solution
exists for ``cold'' initial conditions with near-homogeneous density; the density
in the final state attractor follows a power law, \(\rho \propto |x|^{-\gcrit}\), over 
at least three decades of distance, with \(\gcrit \simeq 0.47\).  The
measured value of the exponent is shallower
than the value \(1/2\) conjectured by \cite{2004MNRAS.350..939B}.
This self-similar state arises
after a period of violent relaxation that redistributes energy while
preserving the energy ranking of the particles. The measurements of the
exponent for the power-law density profile are done at times significantly
after the end of violent relaxation ($t=700$ versus $t=60$), to 
ensure that the coarse-grained density has achieved a stationary form.
Measurements of the exponent obtained
from the \(J(\mu)\) relation can (and must) be done at significantly earlier
times.  These show that
an attractor solution is in place at very early times. However, the
exponent obtained from \(J(\mu)\) is systematically lower than that
obtained from the density at later times. 
The two determinations cannot be compared because at later times 
discreteness effects destroy the phase coherence of the simulations, 
and with it the monotonicity of $J(\mu)$. 
If the initial conditions are warmed by applying smooth velocity perturbations with spatial 
frequencies approaching the Nyquist limit, a cross-over is
reached where self-similarity of the final state is destroyed and 
is replaced by a cored solution.
Cold initial conditions with a steep initial power-law density of
exponent \(\gamma_i > \gcrit\) do not reach the attractor either; they
instead collapse to form a final equilibrium state with
\(\gamma \sim \gamma_i\).

Our simulations provide some insight into the process of violent relaxation and
provide hints that may be useful for the development of an analytical understanding
of the properties of the attractor solution.
For example, we find that many of the scaling laws are in place very early, even during 
the violent relaxation process.  The evolution of some of the scaling relations suggests
that the violent relaxation process finishes first in the outer parts of the structure, 
while the inner particles take longer to redistribute their energies.  
Such insights may then also
be extended to the 3D case, eventually shedding light on the origin of cold dark matter
halo density profiles.

\section*{Acknowledgments}
The authors thank Daniel Holz, Doug Rudd, Mike Warren, and Nadia Zakamska for inspiring
and helpful conversations. A.E.S. was supported in part by the Corning
Glassworks fellowship and the National Science Foundation.


\bibliographystyle{mnras}
\bibliography{odg}

\end{document}